\documentclass[11pt]{article}
\pdfoutput=1

\usepackage{amsmath}
\usepackage{amssymb}
\usepackage{amsthm}
\usepackage{booktabs}
\usepackage{cite}
\usepackage{fancyhdr}
\usepackage[headsep=11pt,headheight=14pt]{geometry}
\usepackage{graphicx}
\usepackage{hyperref}

\title{Formal Semantics of Architectural Decision Models}

\author{Marcin Szlenk\\
Institute of Control and Computation Engineering,\\
Warsaw University of Technology, Nowowiejska 15/19,\\
Warszawa, 00-665, Poland\\
m.szlenk@elka.pw.edu.pl\\
\url{http://www.ia.pw.edu.pl/~mszlenk}}

\date{\today}

\theoremstyle{definition}
\newtheorem{definition}{Definition}

\newcommand{\ADIssue}{\mathit{Issue}}
\newcommand{\ADAlternative}{\mathit{Alternative}}
\newcommand{\M}{\mathcal{M}}
\newcommand{\issues}{\mathit{issues}}
\newcommand{\alternatives}{\mathit{alternatives}}
\newcommand{\issueFor}{\mathit{issueFor}}
\newcommand{\alternativesTo}{\mathit{alternativesTo}}
\newcommand{\compatibleWith}{\mathit{compatibleWith}}
\newcommand{\incompatibleWith}{\mathit{incompatibleWith}}
\newcommand{\forcedBy}{\mathit{forcedBy}}
\newcommand{\triggeredBy}{\mathit{triggeredBy}}
\newcommand{\fundef}{\mathit{def}}
\newcommand{\Power}{\mathcal{P}}

\newcommand{\dom}{\mathbf{dom}}
\newcommand{\range}{\mathbf{rng}}
\newcommand{\Model}{\mathit{Model}}
\newcommand{\Design}{\mathit{Design}}
\newcommand{\conforms}{\mathit{conforms}}
\newcommand{\meaningOf}{\mathit{meaningOf}}

\begin{document}

\maketitle

\begin{abstract}
A software architecture is the result of multiple decisions made
by a software architect. These decisions are called architectural
decisions, as they bring solutions to architectural problems.
Relations between decisions can be captured in architectural
decision models. Such models are then a form of reusable knowledge
for software architects. Several models have been described in the
literature, introducing necessary concepts and relations. These
concepts and relations were usually explained using natural language.
Not much work has been done so far on their formal definitions.
Specifically, such a definition of an architectural decision model
is still missing. The purpose of this paper is filling this gap by
providing the formal definition of an architectural decision model
at both syntax and semantics levels. At the syntax level, different
concepts and relations that are elements of a model have been
mathematically defined. At the semantics level, the meaning of
a model has been defined in a form of \emph{denotational semantics}.
The formalization not only allows for better understanding of
architectural decision models but opens the possibility to reason on
such models, e.g., checking their consistency -- something that is
very limited for the models proposed so far. A practical example of
the semantics of an architectural decision model is also presented.

\end{abstract}

\section{Introduction}

A software architecture is developed as a result of numerous,
sometimes strictly related, decisions. Any decision that is pertaining
to a software architecture is called an \emph{architectural decision}.
The architecture itself, these decisions and the context of these
decisions create the complete architectural documentation.

Several models of architectural decisions have been proposed in the
literature by defining basic concepts and relations in an informal
way~\cite{bosch,tyree,harrison,capilla,jansen,zalewski,kruchten}.
The structural dependencies between concepts and relations were
usually shown with UML class diagrams. Even if they are presented in
a form of mathematical relations with more detailed
constraints~\cite{zimmermann}, only the metamodel (i.e., the abstract
syntax of the language) for documenting architectural decisions is
defined, while the definition of the semantics of its elements is
still missing. This way, properties, like semantical consistency of
elements, cannot be decided. As a consequence, it reduces models'
verifiability. In this paper, the syntax and semantics of an
architectural decision model is formally defined. The goal of this
formalization is twofold. First, it helps in understanding the meaning
of architectural decision models and relations that appear in those
models. Second, it increases the verifiability of architectural
decision models and can be a basis for more advanced software tools
supporting architectural decision modeling.

The remaining part of the paper is organized as follows.
In Section~\ref{RELATED:SEC}, the related work on modeling
architectural decisions is discussed. Basic concepts that
are used in architectural decision models are informally described
in Section~\ref{OVERVIEW:SEC}. Next, the formal syntax and
semantics of an architectural decision model are defined respectively
in Section~\ref{MODEL:SEC} and Section~\ref{SEMANTICS:SEC}. An example
model and its semantics are presented in Section~\ref{EXAMPLE:SEC}.
The conclusions with plans for further work are outlined in
Section~\ref{CONCLUSIONS:SEC}.

\section{Related Work}\label{RELATED:SEC}

Various methods and tools for architectural decision modeling have
been surveyed in~\cite{shahin,tang}. In general, the role of
architectural decision models can be twofold. They can be used to
document decisions that have already been made and/or to ontologize
the decision making process in a given domain. In the first case,
decisions are usually documented in a form of textual descriptions
of their attributes, sometimes accompanied with illustrating
diagrams~\cite{bosch,tyree,harrison,capilla,jansen,zalewski}.
In the latter case, architectural decision models focus on showing
possible decisions and relationships between them within the decision
making process~\cite{kruchten,zimmermann}.

In~\cite{kruchten}, \emph{architectural knowledge} is defined as
architecture design with design decisions, assumptions, context and
other factors that determine a particular solution. A design decision
comprises both a design problem and its solution. The proposed
classification of design decisions covers \emph{existence},
\emph{non-existence}, \emph{property} and \emph{executive} decisions.
Each decision is further characterized with attributes, such as
\emph{scope}, \emph{rationale} or \emph{state}. The possible decision
states are, e.g., \emph{idea}, \emph{tentative}, \emph{decided},
\emph{approved} and \emph{rejected}. Allowed transitions between
states are defined using the state machine. Kinds of relationships
between decisions include, e.g., \emph{enables} (one decision makes
possible the other one), \emph{subsumes} (one decision is wider
than the other one) and \emph{conflicts with} (two decisions
are mutually exclusive), but their semantics is not formally defined.

In~\cite{zimmermann}, decisions made are distinguished from decisions
required from the architect. The proposed UML metamodel for capturing
architectural decisions contains, among others, three core entities:
\emph{ADIssue}, \emph{ADAlternative} and \emph{ADOutcome}.
An instance of ADIssue (an architectural decision issue) informs the
architect that a single architecture design problem has to be solved,
whereas instances of ADAlternative (architectural decision
alternatives) present possible solutions to this problem. These two
entities provide reusable background information. Finally, instances
of ADOutcome (architectural decision outcomes) present an actual
decision made to solve the problem including its rationale. The
mentioned UML metamodel is complemented with formal definitions for a
rich set of relationships between alternatives, alternatives and
issues, and issues alone. However, the formal definitions of their
semantics are missing.

An attempt to formally define the semantics of the elements of
architectural decision models is presented in~\cite{sacha}. It borrows
the notation from~\cite{zimmermann}, introducing sets of
architectural decision alternatives, issues and outcomes. The
proposed semantics defines the meaning of an architectural decision 
as a set of software systems in which this decision is implemented.
Similarly, the meaning of various relationships between issues and
alternatives is defined with reference to respective sets of software
systems. The set of all software systems is finite~\cite{sacha} but
unbounded, what makes the presented definitions suitable for
verifications with the use of \emph{theorem proving} approach, rather
than \emph{model checking}~\cite{jackson}. Another drawback
of this work is the lack of definition of the semantics of the
whole architectural decision model. Each element on the model is
considered separately.

In Section~\ref{OVERVIEW:SEC}, selected concepts and relationships
from the architectural decision models in~\cite{zimmermann} and
\cite{sacha} are presented informally. They will be the starting point
for further formalization of an architectural decision model.

\section{Overview of Basic Concepts}\label{OVERVIEW:SEC}

A simple example of an \emph{architectural decision issue} is the
programming language for implementing a software system. This issue
represents a design concern. The set of \emph{architectural decision
alternatives} for this issue comprises C, Java, Swift, etc. So the
alternatives represent solutions to the issue, and making a decision
on the issue means selecting one of its alternatives. Another example
of an architectural decision issue can be the version of Java
language. The set of alternatives for this issue comprises Java~6,
Java~7, Java~8, etc.

A software architecture is the result of a decision making process.
The set of issues to be resolved is not constant throughout this
process. New issues can be \emph{triggered} (created) by earlier
decisions, in the sense that making one decision leads an architect
to another problem that needs to be resolved. For example, choosing
Java as the programming language triggers the Java version issue.
In this case, making decision on the triggered issue describes
the previous decision more precisely. Choosing a version of Java makes
the general choice of Java more specific. It is worth noting that
if one of the other programming languages was chosen, there wouldn't
be any reason to make a decision on the Java version issue. Choosing
an alternative can trigger many issues and vice versa, an issue can
be triggered by many alternatives.

Two alternatives are \emph{compatible} if both of them can be selected
as the solutions to their respective issues. In other words, such
alternatives work with each other. If it does not happen, they are
said to be \emph{incompatible}. An example of two compatible
alternatives are Java (programming language) and ARM (processor
architecture), as there exists Java VM for ARM processors. An example
of incompatible alternatives are Swift and ARM, because Swift is not
officially supported on ARM processors (at least, at the moment of
writing this paper). The software architecture, where the chosen
programming language is Swift and the chosen processor architecture
is ARM, is simply not implementable.

It has been discussed in~\cite{sacha}, that the compatibility relation
in a set of alternatives is symmetric and reflexive, but not
transitive. Moreover, the relations of compatibility and
incompatibility are complementary: any two alternatives are either
compatible or incompatible, but not both. It is worth noting that
defining two alternatives for the same issue to be compatible or not
is meaningless (e.g., Java and Swift as programming languages). They
are possible solutions to the same problem, so only one of them will
be selected.

\section{Architectural Decision Model}\label{MODEL:SEC}

Below we formally define the abstract syntax of an architectural
decision model. The definition presented clarifies the informal
description from Section~\ref{OVERVIEW:SEC}. The new concern that has
not been discussed before is the \emph{forcing} relation between
alternatives~\cite{sacha}. This relation is defined using the
relations of compatibility and incompatibility. To show the proper
context for the forcing relation, it will be explained in a
footnote.

\begin{definition}[Issues] With $\ADIssue$ we denote a set of all
the architectural decision issues (problems).
\end{definition}

\begin{definition}[Alternatives] With $\ADAlternative$ we denote a
set of all the architectural decision alternatives (possible solutions).
\end{definition}

%
\begin{definition}[Model]\label{MODEL:DEF}
By an \emph{architectural decision model} we understand a tuple
\begin{equation}
\M = (\issues, \alternatives, \issueFor, \compatibleWith, \triggeredBy),
\end{equation}
where:
\begin{enumerate}
\item
$\M.\issues$ is a set of architectural issues:
\begin{equation}
\M.\issues \subseteq \ADIssue.
\end{equation}

\item
$\M.\alternatives$ is a set of alternatives:
\begin{equation}
\M.\alternatives \subseteq \ADAlternative.
\end{equation}

\item
$\M.\issueFor$ is a function of an alternative's issue. The function
maps each alternative to its issue:
\begin{equation}
\M.\issueFor\colon \M.\alternatives \to \M.\issues.
\end{equation}
For the model $\M$, a function of issue alternatives is thus
defined as:\footnote{For a set $A$, $\Power(A)$ denotes the set of all
the subsets of $A$.}
\begin{align}
& \M.\alternativesTo\colon_{\hspace{-0.8ex}\fundef}\ \M.\issues \to
  \Power(\M.\alternatives), \\ \nonumber
& \M.\alternativesTo(i) =_\fundef \\ \nonumber
& \qquad \{\, a \in \M.\alternatives : \M.\issueFor(a) = i \,\}.
\end{align}

\item
$\M.\compatibleWith$ is a function of alternative's compatibility.
The function assigns to each alternative a set of alternatives that
are compatible with it:
\begin{equation}
\M.\compatibleWith\colon \M.\alternatives \to \Power(\M.\alternatives).
\end{equation}
The relation of compatibility is symmetric and reflexive~\cite{sacha}.
Formally:
\begin{align}
& \forall a, a' \in \M.\alternatives \cdot \\ \nonumber
& \qquad a \in \M.\compatibleWith(a') \implies
  a' \in \M.\compatibleWith(a), \\
& \forall a \in \M.\alternatives \cdot a \in \M.\compatibleWith(a).
\end{align}
For the model $\M$, a function of alternative's incompatibility is
thus defined as:\footnote{As it was stated earlier, the relations of
compatibility and incompatibility are complementary. The relation of
incompatibility is here a derived relation that can be computed
from the relation of compatibility.}
\begin{align}
& \M.\incompatibleWith\colon_{\hspace{-0.8ex}\fundef}\ \M.\alternatives
  \to \Power(\M.\alternatives), \\ \nonumber
& \M.\incompatibleWith(a) =_\fundef \\ \nonumber
& \qquad \M.\alternatives \setminus \M.\compatibleWith(a)
\end{align}
and a function of forced alternatives as:\footnote{The definition
presented is based on~\cite{sacha}. Let $a$ be an alternative for
issue $i$ and $a'$ be an alternative for issue $i'$
($i \neq i'$). The alternative $a$ \emph{forces} $a'$ if $a$
is compatible with $a'$ and is incompatible with any other
alternative for $i'$. Such as the incompatibility relation, the
relation of forcing is a derived one, i.e., it can be computed
from other elements of the model.}
\begin{align}
& \M.\forcedBy\colon_{\hspace{-0.8ex}\fundef}\ \M.\alternatives \to
  \Power(\M.\alternatives), \\ \nonumber
& \M.\forcedBy(a) =_\fundef \{\, a' \in \M.\compatibleWith(a) : \\ \nonumber
& \qquad \M.\issueFor(a') \neq \M.\issueFor(a)\ \land \\ \nonumber
& \qquad \M.\alternativesTo(\M.\issueFor(a')) \setminus \{a'\} \subseteq \\ \nonumber
& \qquad \qquad \M.\incompatibleWith(a)\,\}.
\end{align}

\item
$\M.\triggeredBy$ is a function of triggered issues. The function maps
each alternative to a set of new issues created by it: 
\begin{equation}
\M.\triggeredBy\colon \M.\alternatives \to \Power(\M.\issues).
\end{equation}
An alternative cannot create its own issue:
\begin{equation}
\forall a \in \M.\alternatives \cdot
\M.\issueFor(a) \notin \M.\triggeredBy(a).
\end{equation}
\end{enumerate}
\end{definition}
%

\begin{definition}[Models]
With $\Model$ we denote a set of all the architectural decision models
as in definition~\ref{MODEL:DEF}.
\end{definition}

\begin{figure}[bt]
\centering
\includegraphics[scale=0.88]{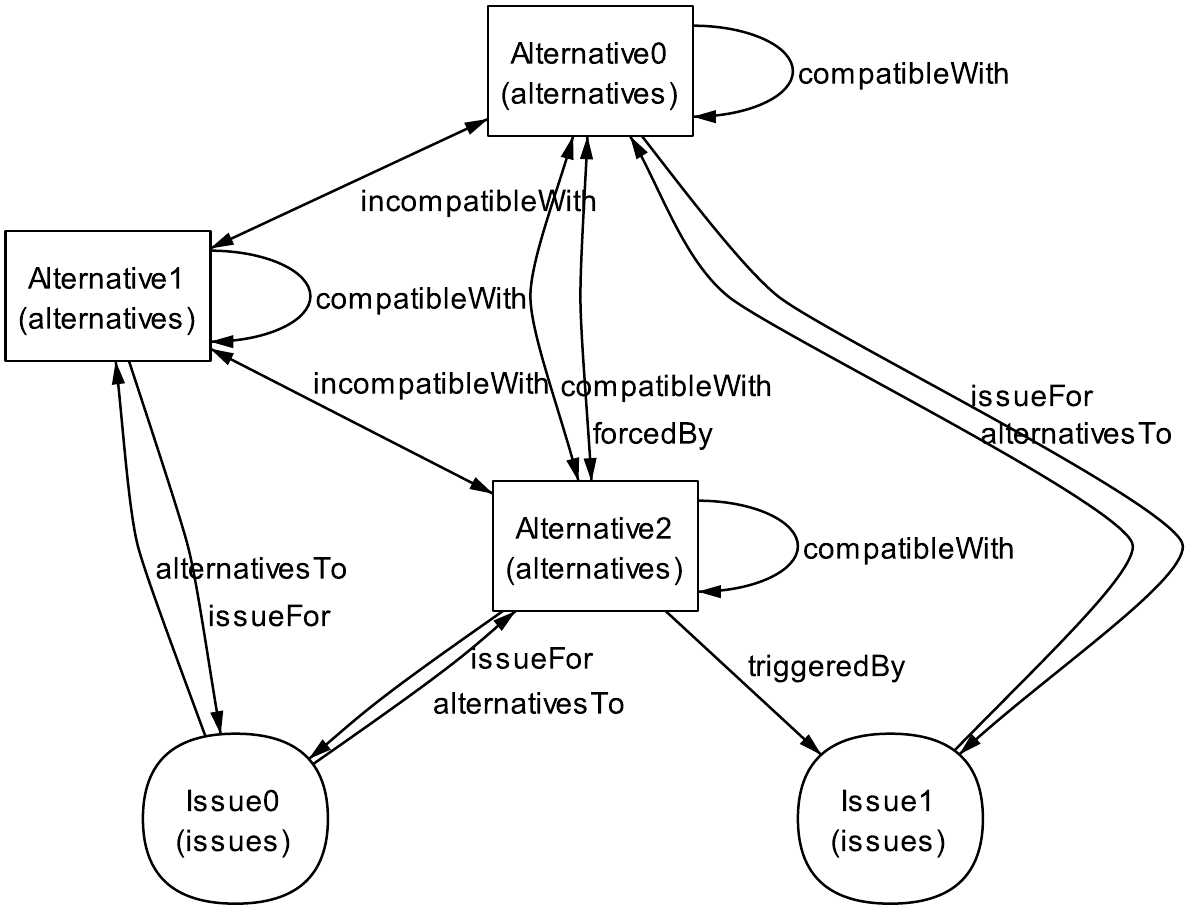}
\caption{An instance of the Alloy model from Appendix~\ref{MODEL:APP}.}
\label{ALLOY:FIG}
\end{figure}

Our past experience with the formalization of UML
models~\cite{szlenk2,szlenk3} has shown that it is highly difficult
to avoid any flaws in such definitions. The kinds of problems we
were struggling with can be summarized with following informal
questions: are all formulas correct? Are they all required formulas?
Are all formulas required? In the current work, to cope with these
problems, the above mathematical definition of an architectural
decision model has been worked out together with its specification in
the Alloy language~\cite{jackson,torlak}. The specification is
presented in Appendix~\ref{MODEL:APP}. The model expressed in Alloy
was being verified with the Alloy Analyzer tool through model
\emph{simulations}, i.e., finding model instances. A number of
instances of the architectural decision model have been generated and
analyzed to assure that all the necessary constraints have been
finally captured and properly expressed. The subject is, however, too
broad to be deeply addressed in this paper. An example instance of the
model, that was obtained with the Alloy Analyzer, is presented in
Fig.~\ref{ALLOY:FIG}.

\section{Semantics of Architectural Decision Model}\label{SEMANTICS:SEC}

Designing a software architecture comprises a set of architectural
decisions. This process can be \emph{driven by} an architectural
decision model that represents architectural knowledge for a specific
technical domain. Such a model carries information about architectural
issues that need to be solved and their possible solutions
(alternatives). It also shows, through the triggering relation, that
some decisions should result from others, and through the
compatibility relation, that some solutions may work or may not work
with each other. Following the information from an architectural
decision model, a software architect decides among alternatives of
subsequent issues. Those decisions (i.e., selected alternatives)
constitute the \emph{design} of a software architecture. Below,
the concepts of a design and its conformity to an architectural
decision model are formally defined.

\begin{definition}[Design]\label{DESIGN:DEF}
A \textit{design} is a partial function
\begin{equation}
d\colon \ADIssue \rightharpoonup \ADAlternative.
\end{equation}
The function maps issues to their solutions, i.e., it defines
alternatives that are selected for architectural decision issues.
\end{definition}

\begin{definition}[Designs]
With $\Design$ we denote a set of all the designs as in
definition~\ref{DESIGN:DEF}.
\end{definition}

%
\begin{definition}[Conformity]
Let $d \in \Design$ and $\M \in \Model$. The design $d$
\emph{conforms to the model} $\M$ and we write
\begin{equation}
\conforms(d, \M),
\end{equation}
if and only if:\footnote{For a function $d$, $\dom(d)$ and $\range(d)$
denote, respectively, the domain and the range of $d$.}
\begin{enumerate}
\item All resolved issues are defined in $\M$:
\begin{equation}
\dom(d) \subseteq \M.\issues.
\end{equation}

\item
A solution to a given issue is one of the alternatives to that issue:
\begin{equation}
\forall i \in \dom(d) \cdot d(i) \in \M.\alternativesTo(i).
\end{equation}

\item
Any two solutions are compatible with each other:
\begin{equation}
\forall i, i' \in \dom(d) \cdot d(i') \in \M.\compatibleWith(d(i)).
\end{equation}

\item
All the top level issues in $\M$ (i.e., not triggered
ones)\footnote{In~\cite{zimmermann}, such issues are called
architectural decision \emph{entry points}.} are resolved, whereas
the lower level issues (i.e., triggered ones) are resolved if, and
only if, the triggering alternative was selected:
\begin{align}
& \forall i \in \M.\issues \cdot i \in \dom(d) \iff \\ \nonumber
& \qquad \big((\nexists a \in \M.\alternatives \cdot
  i \in \M.\triggeredBy(a))\ \lor \\ \nonumber
& \qquad \ \, (\exists a \in \range(d) \cdot i \in \M.\triggeredBy(a))\big).
\end{align}
\end{enumerate}
\end{definition}
%

As it has been mentioned before, an architectural decision model can
represent architectural knowledge for a specific domain. Basing
on this knowledge, one can design the software architecture. All the
designs that conform to this knowledge (i.e., which use given concepts
and respect constraints from a model) can be treated as a
\emph{meaning} of an architectural decision model.

\begin{definition}[Model's meaning]\label{MEANING:DEF}
Let $\M\in \Model$ and $\meaningOf\colon \Model \to \Power(\Design)$
be the function which is defined as:
\begin{equation}
\meaningOf(\M) =_\fundef \{\, d\in \Design : \conforms(d, \M)\,\}.
\end{equation}
The value $\meaningOf(\M)$ refers to the \emph{meaning of} $\M$.
\end{definition}

It may happen that there are no designs conforming to a given model.
In such a situation, the model can be interpreted as
\emph{inconsistent} in a sense that it describes a domain in which one
cannot construct an implementable architecture. A typical example of
inconsistency would be a model forcing an architect to select two
incompatible alternatives. However, due to limited space, that subject
will not be elaborated more here.

\begin{definition}[Model's consistency]\label{CONSISTENT:DEF}
Let $\M \in \Model$. The architectural decision model $\M$ is
\emph{consistent}, if and only if:
\begin{equation}
\meaningOf(\M) \neq \emptyset.
\end{equation}
\end{definition}

\section{Example}\label{EXAMPLE:SEC}

As an example of the semantics of an architectural decision model we
will consider architectural decisions that are being made when
building a robot application in the RAPP system \cite{reppou,szlenk4}.
RAPP is an open-source software platform for developers to create and
deliver robotic applications dedicated to social inclusion of elderly
people. It is the result of a 3-year research project (2013--2016)
funded by the European Commission.

The overall architecture of the RAPP system is composed of two layers:
a RAPP platform and Robot platform. The RAPP platform is located in
the cloud and provides computing-intensive services to robots, such as
machine learning or data mining. It also contains a RAPP store, which
holds RAPP applications (provided by developers) that can be
downloaded and executed on robots. The Robot platform is located on
each robot and allows them to connect to the RAPP platform. It is
responsible for downloading and starting RAPP applications,
calling services provided by the RAPP platform and also offers some
robot-specific services, such as movement control, sound recording
or image capturing.

\begin{figure}[bt]
\centering
\includegraphics[scale=0.88]{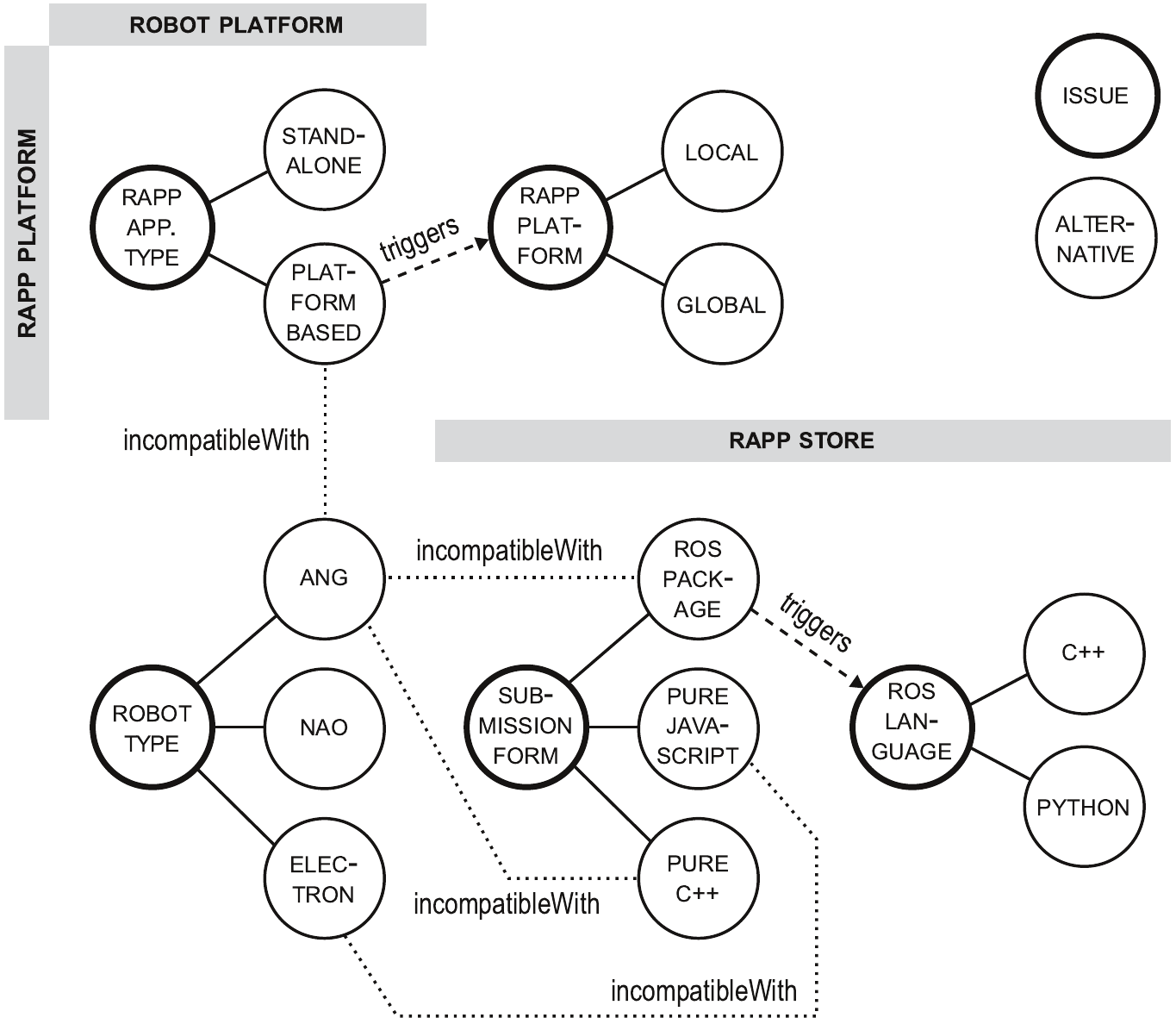}
\caption{Architectural decision model for RAPP. Architectural issues
are grouped into the problems related to: RAPP platform, Robot
platform, and RAPP store. For readability, the compatibility relation
is not shown (it is complementary to the incompatibility relation).}
\label{RAPP:FIG}
\end{figure}

The simplest RAPP applications are the ones using only robot-specific
services of the Robot platform, i.e., not requiring the RAPP
platform. If a developer decides to use services offered by the latter,
he or she can use the global instance of the RAPP platform (being
maintained as part of the RAPP research project) or configure and
use a local instance. At the moment of writing this paper, three
robot types are supported by RAPP: ANG, NAO and Electron, but the
connection to the RAPP platform is fully implemented only for NAO and
Electron. As for holding RAPP applications in the RAPP store, they
can be submitted by developers in one of three forms: as a ROS (Robot
Operating System) package with C\texttt{++} or Python code, pure
JavaScript code, or pure C\texttt{++} code. These possible submission
forms are further restricted by the robot type for which the given
RAPP application has been created. The described architectural
decisions are presented on the model in Fig.~\ref{RAPP:FIG}.

Table~\ref{RAPP:TAB} shows all twenty two of designs (i.e., mappings
of issues to their solutions) that conform to the architectural
decision model presented in Fig.~\ref{RAPP:FIG}. Under
definition~\ref{MEANING:DEF}, they are the meaning of this model.
These conforming designs have been automatically generated with the
Alloy Analyzer, based on the formal definitions proposed in this
paper. The Alloy specification of a design and its conformity is
shown in Appendix~\ref{CONFORMITY:APP}. Under
definition~\ref{CONSISTENT:DEF}, the model considered is certainly
consistent.

\begin{table}[ht]
\caption{The meaning of architectural decision model for RAPP. Each
row presents one design, i.e., the alternatives that are selected for
respective issues (\emph{none} means undefined value).}\label{RAPP:TAB}

\medskip\centering
\begin{tabular}{ccccc} \toprule
RAPP app. type & RAPP platform & Robot type & Submission form & ROS language\\
\midrule
Platform based & Local & NAO & ROS package & C\texttt{++} \\
Platform based & Local & NAO & ROS package & Python \\
Platform based & Local & NAO & Pure JavaScript & \emph{none} \\
Platform based & Local & NAO & Pure C\texttt{++} & \emph{none} \\
Platform based & Local & Electron & ROS package & C\texttt{++} \\
Platform based & Local & Electron & ROS package & Python \\
Platform based & Local & Electron & Pure C\texttt{++} & \emph{none} \\
Platform based & Global & NAO & ROS package & C\texttt{++} \\
Platform based & Global & NAO & ROS package & Python \\
Platform based & Global & NAO & Pure JavaScript & \emph{none} \\
Platform based & Global & NAO & Pure C\texttt{++} & \emph{none} \\
Platform based & Global & Electron & ROS package & C\texttt{++} \\
Platform based & Global & Electron & ROS package & Python \\
Platform based & Global & Electron & Pure C\texttt{++} & \emph{none} \\
Stand-alone & \emph{none} & ANG & Pure JavaScript & \emph{none} \\
Stand-alone & \emph{none} & NAO & ROS package & C\texttt{++} \\
Stand-alone & \emph{none} & NAO & ROS package & Python \\
Stand-alone & \emph{none} & NAO & Pure JavaScript & \emph{none} \\
Stand-alone & \emph{none} & NAO & Pure C\texttt{++} & \emph{none} \\
Stand-alone & \emph{none} & Electron & ROS package & C\texttt{++} \\
Stand-alone & \emph{none} & Electron & ROS package & Python \\
Stand-alone & \emph{none} & Electron & Pure C\texttt{++} & \emph{none} \\
\bottomrule
\end{tabular}
\end{table}

\section{Conclusions}\label{CONCLUSIONS:SEC}

In the paper we have proposed a concise formalization of an
architectural decision model. The main application of the presented
model is to support a decision making process through capturing
reusable information about problems to be solved and their potential
solutions. The set of all designs of the architecture that a
software architect would create using this information is considered
here as the meaning of a given architectural decision model. As can be
seen from the example presented, the semantics defined is relatively
easy to be computed (however, we have not considered the problem 
of computational complexity in the paper). Thanks to this feature,
it can be useful not only for the theory of architectural decision
models but also used practically in software tools for modeling
architectural decisions.

The formalization presented helps in understanding the meaning of
architectural decision models and relations that appear in those
models. It opens possibility to reason on such models in a way
similar to the one proposed for UML models in~\cite{szlenk2,szlenk3}.
However, an issue related to reasoning about consistency of an
architectural decision model has only been briefly touched upon here.
Together with the problem of semantical equivalence of two models, it
is an interesting direction of further theoretical research.

\section*{Acknowledgments}

The author would like to thank Prof. Krzysztof Sacha~\cite{sacha}
for his inspiration and encouragement to currying out this research.

\appendix
\section{Architectural Decision Model in Alloy}\label{MODEL:APP}

\begin{verbatim}
sig Issue {}
sig Alternative {}
one sig M {
    issues: set Issue,
    alternatives: set Alternative,
    issueFor: alternatives -> one issues,
    compatibleWith: alternatives -> alternatives,
    triggeredBy: alternatives -> issues, 
    alternativesTo: issues -> alternatives,
    incompatibleWith: alternatives -> alternatives,
    forcedBy: alternatives -> alternatives
}{
    all i: issues | alternativesTo[i] =
        { a: alternatives | issueFor[a] = i }
    all a, a': alternatives |
        a in compatibleWith[a'] implies a' in compatibleWith[a]
    all a: alternatives | a in compatibleWith[a]
    all a: alternatives | incompatibleWith[a] =
        alternatives - compatibleWith[a]
    all a: alternatives | forcedBy[a] =
        { a': compatibleWith[a] | issueFor[a'] != issueFor[a] and
        (alternativesTo[issueFor[a']] - a') in incompatibleWith[a] }
    all a: alternatives | issueFor[a] not in triggeredBy[a]
}
\end{verbatim}

\section{Design and Conformity in Alloy}\label{CONFORMITY:APP}

\begin{verbatim}
sig Issue { d: lone Alternative }
sig Alternative {}
one sig M { ... }{ ... }
fun dom[f: Issue -> Alternative]: set Issue {
    f.Alternative
}
fun rng[f: Issue -> Alternative]: set Alternative {
    Issue.f
}
pred conformity {
    dom[d] in M.issues
    all i: dom[d] | d[i] in M.alternativesTo[i]
    all i, i': dom[d] | d[i'] in M.compatibleWith[d[i]]
    all i: M.issues | i in dom[d] iff
        ((no a: M.alternatives | i in M.triggeredBy[a]) or
         (some a: rng[d] | i in M.triggeredBy[a]))
}
\end{verbatim}


\begin{thebibliography}{00}

\bibitem{bosch} J. Bosch and A. Jansen, Software Architecture as a Set
of Architectural Design Decisions, in {\it Proc. 5th Working IEEE/IFIP
Conf. on Software Architecture $($WICSA'05$)$}, IEEE Computer Society,
2005, pp.~109--120.

\bibitem{tyree} J. Tyree and A. Akerman, Architecture Decisions:
Demystifying Architecture, {\it IEEE Software} {\bf 22}(2) (2005)
19--27.

\bibitem{harrison}
N. B. Harrison, P. Avgeriou, and U. Zdun, Using Patterns
to Capture Architectural Decisions, {\it IEEE Software} {\bf 24}(4)
(2007) 38--45.

\bibitem{capilla} R. Capilla, F. Nava, and J. C. Due\~{n}as, Modeling
and Documenting the Evolution of Architectural Design Decisions, in
{\it Proc. 2nd Workshop on Sharing and Reusing Architectural Knowledge
-- Architecture, Rationale, and Design Intent $($SHARK/ADI'07$)$},
IEEE Computer Society, 2007, pp.~9--16.

\bibitem{jansen}
A. Jansen, P. Avgeriou, and J. S. van der Ven, Enriching
Software Architecture Documentation, {\it J. Syst. Software}
{\bf 82}(8) (2009) 1232--1248.

\bibitem{zalewski}
A. Zalewski, S. Kijas, and D. Sokolowska, Capturing Architecture
Evolution with Maps of Architectural Decisions 2.0,
in {\it Proc. 5th European Conf. on Software Architecture
$($ECSA 2011$)$}, LNCS 6903, Springer-Verlag, 2011, pp.~83--96.

\bibitem{kruchten} P. Kruchten, P. Lago and H. van Vliet, Building Up
and Reasoning About Architectural Knowledge, in {\it Proc. 2nd Int.
Conf. on the Quality of Software Architectures $($QoSA 2006$)$},
LNCS 4214, Springer-Verlag, 2006, pp.~43--58.

\bibitem{zimmermann}
O. Zimmermann, J. Koehler, F. Leymann, R. Polley, and N. Schuster,
Managing architectural decision models with dependency relations,
integrity constraints, and production rules, {\it J. Syst. Software}
{\bf 82}(8) (2009) 1249--1267.

\bibitem{shahin}
M. Shahin, P. Liang, and M. R. Khayyambashi, Architectural design
decision: Existing models and tools, in {\it Proc. Joint 8th Working
IEEE/IFIP Conf. on Software Architecture \& 3rd European Conf. on
Software Architecture $($WICSA/ECSA$)$}, IEEE Computer Society,
2009, pp.~293--296.

\bibitem{tang}
A. Tang, P. Avgeriou, A. Jansen, R. Capilla, and M. A. Babar,
A comparative study of architecture knowledge management tools,
{\it J. Syst. Software} {\bf 83}(3) (2010) 352--370.

\bibitem{sacha} K. Sacha, On the Semantics of Architectural Decisions,
{\it Int. J. Soft. Eng. Knowl. Eng.} {\bf 26}(02) (2016) 333--346.

\bibitem{szlenk2}
M. Szlenk, Formal Semantics and Reasoning about UML Class Diagram,
in  {\it Proc. Int. Conf. on Dependability of Computer Systems
$($DepCoS-RELCOMEX 2006$)$}, IEEE Computer Society, 2006, pp.~51--58.

\bibitem{szlenk3}
M. Szlenk, UML Static Models in Formal Approach, in {\it
Balancing Agility and Formalism in Software Engineering},
eds B. Meyer, J. R. Nawrocki, and B. Walter, LNCS 5082,
(Springer, Heidelberg, 2008), pp.~129--142.

\bibitem{jackson}
D. Jackson, Software Abstractions: Logic, Language, and Analysis,
(MIT Press, 2012).

\bibitem{torlak}
E. Torlak, M. Taghdiri, G. Dennis, and J. P. Near,
Applications and extensions of Alloy: past, present and future,
{\it Math. Structures Comput. Sci.} {\bf 23}(4) (2013) 915--933.

\bibitem{reppou}
S. E. Reppou, E. G. Tsardoulias, A. M. Kintsakis, A. L. Symeonidis,
P. A. Mitkas, F. E. Psomopoulos, G. T. Karagiannis, C. Zielinski,
V. Prunet, J.-P. Merlet, M. Iturburu, and A. Gkiokas,
RAPP: A Robotic-Oriented Ecosystem for Delivering Smart User
Empowering Applications for Older People, {\it Int. J. Soc. Robot.}
{\bf 8}(4) (2016) 539--552.

\bibitem{szlenk4}
M. Szlenk, C. Zielinski, M. Figat, and T. Kornuta,
Reconfigurable Agent Architecture for Robots Utilising Cloud Computing,
in {\it Progress in Automation, Robotics and Measuring Techniques.
Vol. 2 Robotics}, eds R. Szewczyk, C. Zielinski, and M. Kaliczynska,
AISC 351, (Springer, 2015), pp.~253--264.
\end{thebibliography}
\end{document}